\documentclass[twocolumn,english]{extarticle}
\usepackage{mathptmx}

\usepackage[T1]{fontenc}
\usepackage[latin9]{inputenc}
\usepackage{geometry}
\usepackage{textcomp}
\usepackage{algorithm2e}
\usepackage{amsthm}
\usepackage{amsmath}
\usepackage{graphicx}
\usepackage{amssymb}
\usepackage{graphicx}
\usepackage{url}
\usepackage{graphicx,amssymb,float,makeidx,amsmath,pstricks,epsf}
\usepackage{cite}
\usepackage{geometry}
\makeatletter

\newcommand{\lyxaddress}[1]{
\par {\raggedright #1
\vspace{1.4em}
\noindent\par}
}

\usepackage{fancyhdr}          
\date{}

\makeatother

\usepackage{babel}
\begin{document}

\title{\textbf{A Novel Mataheuristic based Interference Alignment for $K$-User Interference Channel : A Comparative Study}}

\author{Lysa AIT MESSAOUD$^{1}$, Fatiha MERAZKA$^{1}$ and Daniel MASSICOTTE$^{2}$}

\maketitle

\lyxaddress{$^{1}$LISIC Lab. Telecommunication department, Electronic and Computer Science Faculty,USTHB University, BP. 32, EL-Alia Bab Ezzouar 16111, Algiers, Algeria.}

\lyxaddress{$^{2}$ Université du Québec à Trois-Rivières, Trois-Rivières, Canada}

\begin{abstract}
This paper presents a new Interference Alignment (IA) scheme for $K$-User Multiple Input Multiple Output (MIMO) Interference Channel (IC) based on two metaheuristics, namely Particle Swarm Optimization (PSO) and Artificial Bee Colony (ABC) Algorithm. Tackling interference is an essential issue in wireless communications to which Interference Alignment (IA) provides a promising solution.  However, IA still lacks of explicit and straightforward design procedures. In fact, most of IA procedures aim to minimize a certain Interference Leakage (IL) which measures the effect of the interference on the network, this results in complex optimization tasks involving a large amount of decision variables, together with a problem of convergence of the IA solutions. In this paper the IA optimization is performed using PSO, ABC and their cooperative counterparts, more suitable for large scale optimization. A comparison between the four algorithms is also carried out. The cooperative proposed approaches seem to be promising.
\end{abstract}
\section{Introduction}
The continued development in wireless communications requires an increasingly high transmission rate; this is achieved especially through the multiantenna technology also called Multiple-Input Multiple-Output (MIMO) technology \cite{MIMO}. Since their advent in the late 1990s, MIMO systems are more and more imperative for improving throughput and/or reliability in wireless communications. Indeed, the transmission rate increases with the number of antennas, without increasing the bandwidth which remains an expensive and limited resource \cite{MMIMO}.

More particularly, a $K$-User MIMO Interference Channel (MIMO IC) is constituted of $K$ transmitter/receiver pairs, every transmitter communicates multiple data streams to its corresponding receiver generating interference at all other receivers. The mitigation of this interference is still a major concern to which Interference Alignment (IA) provides a promising solution \cite{IA}. IA aligns, at each receiver, the interference in the smallest possible part of the space formed by the available signaling dimensions; the remaining part will contain the useful signal that will be recovered using a decoding matrix. Reference \cite{survey2016} offers a recent and complete state-of-art about IA.

IA solutions are mainly iterative as closed form solutions are only possible for certain low-dimensional configurations of MIMO IC \cite{IA}. IA iterative schemes result in complex optimization tasks involving a large amount of decision variables, together with a problem of convergence. If this convergence is not reached, only suboptimal solutions can be found at the expense of high computational complexity that increases rapidly with the size of the MIMO IC (number of users, antennas or data streams). Reference \cite{CS} contains a comparative study of usual optimization algorithms dedicated to $K$-User MIMO IC AI. It is clearly shown that finding effective and straightforward IA solution remains a challenging problem, especially as the wireless networks data traffic is increasing significantly and continuously, i.e. the IA optimization problem expands in scale and thereby in complexity.

On the other hand, Particle Swarm Optimization (PSO) \cite{PSO}, and Artificial Bee Colony (ABC) Algorithm \cite{ABC} are the most popular stochastic optimization tools. PSO and ABC imitate the social behavior of swarms in foraging for food, known for combining simplicity and efficiency. PSO and ABC were effectively applied in a large variety of engineering problems. 

In this paper, we propose to implement these two metaheuristics to achieve IA for $K$-User MIMO IC. Given that the optimization problem specific to the IA is of large scale. It would be better to use more appropriate approaches like Cooperative Coevolution (CC). In fact, CC is a framework proposed by Potter in \cite{CC} to alleviate the weakness of the most stochastic optimization algorithms when dealing with high-dimensional search spaces. In \cite{CPSO} and \cite{CABC}, authors applied Potter's technique to the PSO and ABC respectively, leading to a new cooperative model for each of the two metaheuristics. In this paper, PSO and ABC algorithms and their respective cooperative counterparts,CPSO and CABC, are applied to achieve IA.

This paper is organized as follows: Section II states the IA problem. Section III provides an overview of PSO, ABC and CC algorithms. Simulation results are presented in Section IV. The paper is concluded in Section V. 
\section{Problem Statement}
\begin{figure}[ht]
\centering
\includegraphics[scale=0.4]{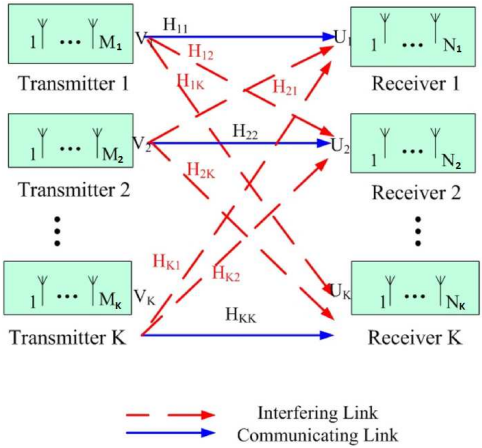}
\caption{$K$-User MIMO IC representation.}
\label{fig:kusermimoIC}
\end{figure}

As depicted in Fig.\ref{fig:kusermimoIC}, a $K$-User MIMO IC consists of $K$ transmitter-receiver pairs equipped each with $M_i$ and $N_i$ antennas, $i=1, \cdots,K$, respectively. $d_i$ is the number of data streams sent by the $k$-th transmitter to its corresponding receiver. Like in \cite{P3}, this system is expressed as $\prod_{i=1}^{K}\left(M_i \times N_i,d_i \right)$. The received signal is given by
\begin{equation}
\begin{aligned}
\mathbf{z}_i=& \mathbf{U}_i^H\mathbf{H}_{ii}\mathbf{V}_i\mathbf{s}_i+\sum_{i\neq j}\mathbf{U}_i^H\mathbf{H}_{ij}\mathbf{V}_j\mathbf{s}_j+\mathbf{n}_i,\\
& i=1, \cdots,K
\end{aligned}
\end{equation}
where $\mathbf{U}_i\in \mathbb{C}^{N_i \times d_i}$ and $\mathbf{V}_i\in \mathbb{C}^{M_i \times d_i}$ are the decoding and precoding matrices, respectively; $\mathbf{H}_{ij} \in \mathbb{C}^{N_i \times M_j}$ is the channel coefficients matrix between transmitter $j$ and receiver $i$, $\mathbf{s}_i \in \mathbb{C}^{d_i}$ are the symbols transmitted by user $i$ and $\mathbf{n}_i \in \mathbb{C}^{d_i}$ is the additive white Gaussian noise at the $i$th receiver. In order to perform IA, the decoding and precoding matrices must be calculated so as to fulfill the following equations \cite{P3}
\begin{equation}
\mathbf{U}_i^H\mathbf{H}_{ij}\mathbf{V}_j=0,~\forall i\neq j
\label{eq:P3.2}	
\end{equation}
\begin{equation}
\text{rank} \left(\mathbf{U}_i^H\mathbf{H}_{ii}\mathbf{V}_i\right)=d_i,~\forall i.
\label{eq:P3.3}	
\end{equation}

Note that if the matrices $\mathbf{U}_i$ and $\mathbf{V}_j$ are full column rank and if the channel matrices $\mathbf{H}_{ij}$ do not have any special structure, condition (\ref{eq:P3.3}) is almost surely satisfied \cite{P3}. This is verified in the calculations performed in this study, since the channel matrices $\mathbf{H}_{ij}$ are generated randomly.

Hence, the problem is confined to verify condition (\ref{eq:P3.2}). For this, it is proposed in \cite{P3} to optimize the variables in $\mathbf{U}_i$ and $\mathbf{V}_j$ concatenated into a single vector $\mathbf{x}=[\text{vec}\left(\mathbf{V}_1\right)^T,\cdots,\text{vec}\left(\mathbf{V}_K\right)^T,\text{vec}\left(\mathbf{U}_1^H\right)^T, \cdots,$ $ \text{vec}\left(\mathbf{U}_K^H\right)^T]^T$, where $\text{vec}(\mathbf{A})$ denotes the vector obtained by stacking the columns of matrix $\mathbf{A}$ below one another. Accordingly of definitions above, $\mathbf{x}$ consists of $N_v=\sum_i(M_i+N_i)d_i$ variables.

Define as $\mathbf{r}(\mathbf{x})$ the function evaluating the residuals of the equations in (\ref{eq:P3.2}) which consists of $N_e=\sum_{i\neq j}d_id_j$ scalar equations, i.e., $\mathbf{r}(\mathbf{x})=[\mathbf{r}_ {21}^T, \cdots, \mathbf{r}_ {(K-1)K}^T]^T$, where $\mathbf{r}_{ij}=\text{vec}(\mathbf{U}_i^H\mathbf{H}_{ij}\mathbf{V}_j)$, to be feasible, the system $\mathbf{r}:\mathbb{C}^{N_v} \rightarrow \mathbb{C}^{N_e}$ must verify $N_v \geq  N_e$. Finally, in order to obtain a mono objective optimization problem, authors in \cite{P3} express the cost function, also called Interference Leakage (IL), as
\begin{equation}
f(\mathbf{x})=\mathbf{r}(\mathbf{x})^H\mathbf{r}(\mathbf{x}):	\mathbb{C}^{N_v}\rightarrow \mathbb{R}
\label{eq:IL}	
\end{equation}
\section{Proposed Solution}
PSO and ABC are population-based optimization algorithms. Within the population, referred to \emph{swarm}, each particle is candidate to be a solution to the optimization problem. At each iteration, every particle moves in the search space toward a better position (solution) depending on the value of its fitness and the information provided by the other particles of the population. Let $S\subseteq \mathbb{R}^n$ an $n$-dimensional search space such that $n$ is the number of decision variables, in a swarm containing $N$ particles, the position of the $i$-th particle is an $n$-dimensional vector $x_i=(x_{i1},x_{i2},\cdots,x_{in})\in S$.
\subsection{PSO algorithm}
PSO was introduced by Kennedy and Eberhart in 1995 \cite{PSO}, in this algorithm each particle modifies its velocity, $v=(v_{i1},v_{i2},\cdots,v_{in})\in S$, following its best position achieved so far, $p_i=(p_{i1},p_{i2},\cdots,p_{in})\in S$, and the global best position of the population. If $g$ is the index of the particle that attained the best previous position among the entire swarm, and $t$ the iteration counter, particle velocity and position are respectively updated as \cite{PSO1}
\begin{equation}
\begin{aligned}
v_{id}(t+1) = & \omega |p_{i^{'}d}(t)-p_{id}(t)|\text{sign}(v_{id}(t)) \\
              & +r(p_{id}(t)-x_{id}(t))+(1-r)(p_{gd}(t)-x_{id}(t))\\      
\end{aligned}
\label{eq:PSAVPSO.8}	
\end{equation}
\begin{equation}
x_{id}(t+1) = x_{id}(t) + v_{id}(t+1)
\label{eq:PSAVPSO.9}	
\end{equation}

where $i = 1, 2, \cdots N$ is the particle's index, $d = 1, 2,\cdots, n$ indicates the particle's $d$-th component, $r \in U[0,1]$, $i^{'}\in intU[0,1]$, $\omega$ is a scaling parameter, and $\text{sign}(v_{id}(t))$ is the sign of $v_{id}(t)$. The velocity update equation (\ref{eq:PSAVPSO.8}) is slightly different from that usually used by PSO. This improved update rule highlights the exploration and exploitation abilities of the particles, which are adjusted with only one parameter, $\omega$.If $\omega > 1$, the velocity $\omega |p_{i^{'}d}(t)-p_{id}(t)|$ increases which expands the search scope of the swarm, this improves exploration ability of the swarm but reduces its convergence speed. If $\omega < 1$, the velocity $\omega |p_{i^{'}d}(t)-p_{id}(t)|$ decreases, which shrinks the search scope of the swarm, this improves the exploitation ability of the swarm but increases its convergence speed; thereby the swarm can be trapped into local optimum. To get a good balance between exploitation and exploration, it would be better to take $\omega =1$. Or alternatively, $\omega$ may be set to $\omega = cr_3$, where $c$ is a parameter and $r_3 \in U[0,1]$. Note that if $c = 2$, then $\omega = 2r3$, and the mean value of $\omega$ is 1; if $c < 2$, then the mean value of $\omega < 1$. The PSO algorithm is summarized in Table \ref{A1}.

\begin{flushleft}
\begin{table}
\begin{tabular}{l}
\hline
\textbf{Algorithm 1} PSO Algorithm\\
\hline
\textbf{1:}~Initialize Population\\
\textbf{2:}~\textbf{repeat}\\
\textbf{3:}~~~~~~~Calculate fitness values of particles\\
\textbf{4:}~~~~~~~Modify the best particles in the swarm\\
\textbf{5:}~~~~~~~Choose the best particle\\
\textbf{6:}~~~~~~~Calculate the velocities of particles\\
\textbf{7:}~~~~~~~Update the particle positions\\
\textbf{8:}~\textbf{until} requirements are met\\
\hline
\end{tabular}
\caption{Pseudocode for PSO algorithm \cite{ABC1}.}
\label{A1}
\end{table} 
\end{flushleft}
\subsection{ABC algorithm}
Introduced by Karaboga in 2005 \cite{ABC}, ABC imitates the foraging behavior of a honeybee colony. The population of artificial bees is divided into two groups, the \emph{employed bees} and the unemployed bees (the \emph{onlookers} and \emph{scouts}). The number, $SN$, of the employed bees or the onlooker bees is equal to the number of solutions (food source positions) in the population. At each optimization cycle, employed bees search the food around the food sources, and then inform the onlooker bees about the quality of these sources. The onlookers select the good sources according to the received information and further search the foods around. The employed bees that abandon, after several tries, their unpromising food sources to investigate new ones become scout bees.The new food sources position are obtained using the following equation \cite{ABC1}

\begin{equation}
v_{ij}=x_{ij}+\varphi_{ij}.(x_{ij}-x_{kj})
\label{eq:CABC.1}	
\end{equation}

for $j\in \left\{1 \cdots n\right\}$ where $n$ is the number of dimensions, $\varphi_{ij}$ is a random number uniformly distributed in the range [-1,1], $k$ is the index of a randomly chosen solution, and $x$ and $v$ are the current and updated solutions respectively. 

Is associated with each food source $i$, $i={1,2,\cdots,SN}$, a probability given by
\begin{equation}
p_i=\frac{fit_i}{\sum_{j=1}^{SN}fit_j}
\label{eq:CABC.2}	
\end{equation}
where $fit_i$ is the fitness of the $i^{\text{th}}$ food source. Every onlooker bee chooses a random food source according to the probability (\ref{eq:CABC.2}), then  tries to find a better food source around the selected one using equation (\ref{eq:CABC.1}). If a food source cannot be improved for a predetermined number of cycles, referred to as $Limit$, this food source is abandoned. The employed bee that was exploiting this food source becomes a scout that looks for a new food source by randomly searching the problem domain. Table \ref{A2} shows the ABC algorithm, the implemented algorithm for the	realization of this study is given in detail in \cite{ABC2}, we have omitted to include it for lack of space.
\begin{flushleft}
\begin{table}
\begin{tabular}{l}
\hline
\textbf{Algorithm 2} ABC Algorithm\\
\hline
\textbf{1:}~Initialize the population of solutions $x_i=(x_{i1},x_{i2},\cdots,x_{in})$,\\$i=1, \cdots, SN$\\
\textbf{2:}~Evaluate the population\\
\textbf{3:}~$cycle=1$\\
\textbf{4:}~\textbf{repeat}\\
\textbf{5:}~~~~~~Produce new solutions $v_i$ for the employed bees \\~~~~~~~~~~by using (\ref{eq:CABC.1}) and evaluate them\\
\textbf{6:}~~~~~~Apply the greedy selection process for the employed bees\\
\textbf{7:}~~~~~~Calculate the probability values $p_i$ for the solutions $x_i$ by\\~~~~~~~~~~(\ref{eq:CABC.2})\\
\textbf{8:}~~~~~~Produce the new solutions $v_i$ for the onlookers from the \\~~~~~~~~~~solutions $x_i$ selected depending on $p_i$ and evaluate them\\
\textbf{9:}~~~~~~Apply the greedy selection process for the onlookers\\
\textbf{10:}~~~~Determine the abandoned solution for the scout, if exists,\\~~~~~~~~~~and replace it with a new randomly produced solution\\
\textbf{11:}~~~~Memorize the best solution achieved so far\\
\textbf{12:}~~~~$cycle=cycle+1$\\
\textbf{13:}~~~~\textbf{until} $cycle = MCN$, $MCN$ is the total number of cycles\\
\hline
\end{tabular}
\caption{Pseudocode for ABC algorithm \cite{ABC1}.}
\label{A2}
\end{table} 
\end{flushleft}
\subsection{CC algorithm}
Cooperative Coevolution (CC) is a promising framework for dealing with large scale optimization problems. Based on multiple agent principle, CC algorithm suggest to divide the high-dimensional search space by splitting the solution vectors into smaller vectors, then each of these smaller solution vectors is optimized by a separate mechanism (here PSO and ABC algorithms)\cite{CC}.

Subcomponents sizes depend on whether the problem is separable or not, a separable problem deals with non-interacting variables which can be optimized as separate problems of lower dimensionality, while a nonseparable problem deals with interacting variables which must be grouped together and optimized jointly. Since the interdependence of decision variables in AI problems is still not analyzed, we opt in this study for the simplest form of the CC algorithm which adopts equally 1-D (i.e. one dimension) sized subcomponents. In other words, each decision variable will be optimized by its own optimization process. Cooperative PSO (CPSO) and Cooperative ABC (CABC) are widely detailed in \cite{CPSO} and \cite{CABC} respectively.

Remain to mention that CPSO and CABC use a population of $n$-dimensional vectors, these vectors are divided into $n$ swarms of 1-D vectors (i.e. in $n$ 1-D optimization problems. That is, each swarm aims to optimize a single component (dimension) of the original solution vector.Whereas each decision variable is optimized separately, the objective function evaluation requires an $n$-dimensional vector as input. The original $n$-dimensional vector must be recovered by taking the global best particle from each of the $n$ swarms and concatenating them to form what is called a \emph{context vector} which will be used to evaluate the objective function \cite{CC,CPSO,CABC}.
\section{Simulation results}
To show the interest of the proposed approach, PSO, CPSO, ABC and CABC algorithms are applied to three $K$-User MIMO IC scenarios. The transmitters and receivers are equipped with $M=N=5$ antennas each, and every transmitter aims to send $d=2$ data streams to its corresponding receiver. These settings chosen according to our main reference \cite{P3} allow a comparison between the two approaches. This scenario is tested with an increasing $K$ of 3, 7 and 13. The three scenarios are listed in the tables \ref{resultsPSO} and \ref{resultsABC} with their respective dimensions (number of decision variables).
For each scenario, the number of decision variables is equal to the total number of complex elements in $\mathbf{U}$ and $\mathbf{V}$ matrices, this sum is equal to $(K \times N \times d)+(K \times M \times d)$. 

However, the fitness $f(\mathbf{x})$ given by equation (\ref{eq:IL}) is a real-valued function of complex valued variables, i.e. not homomorphic \cite{wirtinger}, the optimization of this kind of functions uses the so-called Wirtinger Calculus \cite{wirtinger2}. Simply put, when optimizing real functions of one or more complex variables, we consider each complex variable as two real independent variables, the real part and the imaginary part. Thus optimization can be done as for multidimensional real functions \cite{wirtinger}. According to this, the total number of the decision variables involved in our optimization process is equal to $2 \times (K \times N \times d)+2\times (K \times M \times d)$.

the entries of the MIMO channels are independent and identically distributed complex Gaussian variables with zero mean and unit variance, and the calculations were performed under MATLAB R2013a software. For each algorithm, several test runs was executed before fixing each parameter, afterward each scenario was optimized over 10 independents runs.

\begin{figure}  
\begin{flushleft}
\includegraphics[width=0.45\textwidth]{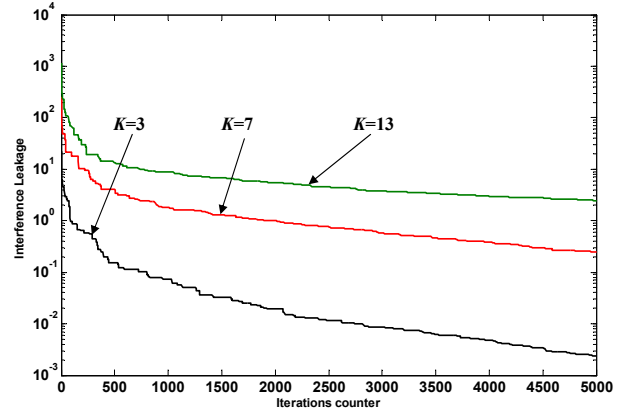}
  \caption{PSO simulation results for the three scenarios.}
  \label{fig:PSO}
\end{flushleft}
\end{figure}

\begin{figure} 
\begin{flushleft}
\includegraphics[width=0.45\textwidth]{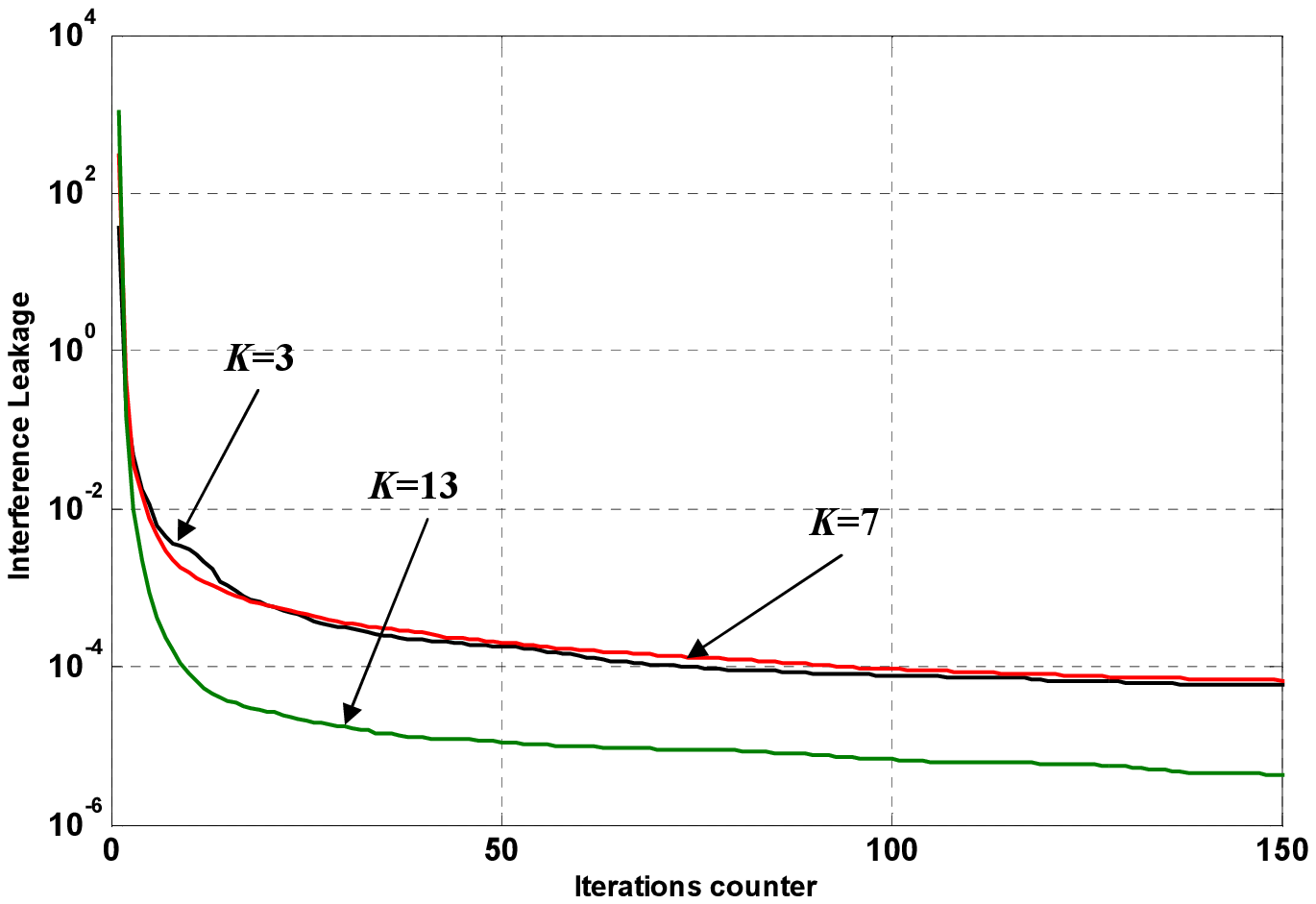}
  \caption{CPSO imulation results for the three scenarios.}
  \label{fig:CPSO}
\end{flushleft}
\end{figure}

Firstly, PSO algorithm was run with $\omega=3$ and a swarm size of 100. Secondly, the CPSO algorithm was run with $\omega=10^-3$ and as many swarms as decision variables, each swarm consists of 50 particles. Figure \ref{fig:PSO} and Figure \ref{fig:CPSO} show the results of simulations (evolution of the Interference Leakage with the iterations counter) for PSO and CPSO respectively. For PSO the slow convergence is important, and the calculation was stopped intentionally after 5000 iterations. As for the CPSO, the convergence showed a much better behavior. According to Table \ref{resultsPSO}, the IL has reached satisfactory values, approximately equal to the IL average obtained in reference \cite{P3} (of about $10^{-5}$).

\begin{figure}  
\begin{flushleft}
\includegraphics[width=0.4\textwidth]{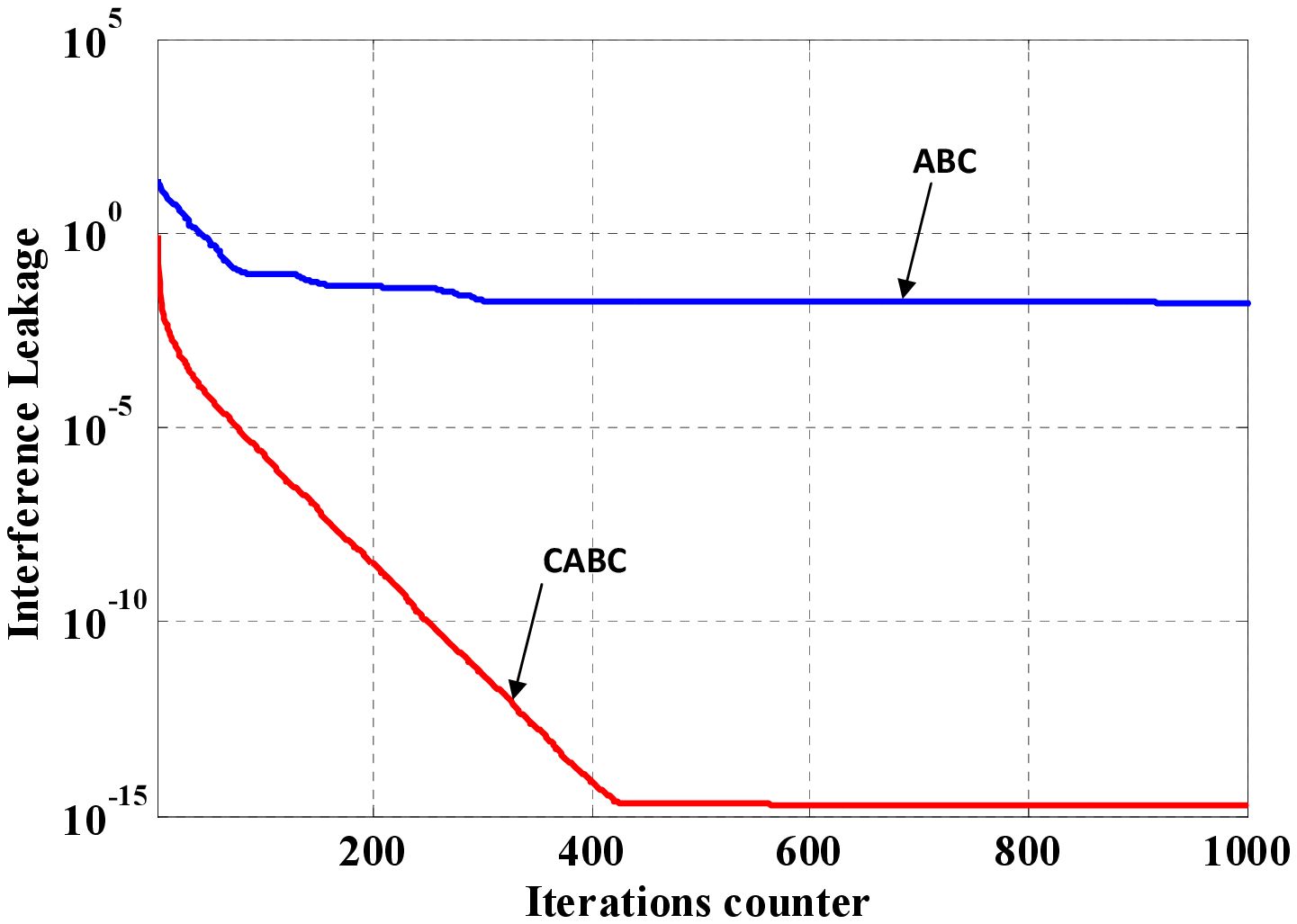} 
 \caption{ABC and CABC simulation results for $\left(5 \times 5, 2\right)^{3}$ scenario.}
  \label{fig:fig3}
\end{flushleft}
\end{figure}

\begin{figure}  
\begin{flushleft}
\includegraphics[width=0.4\textwidth]{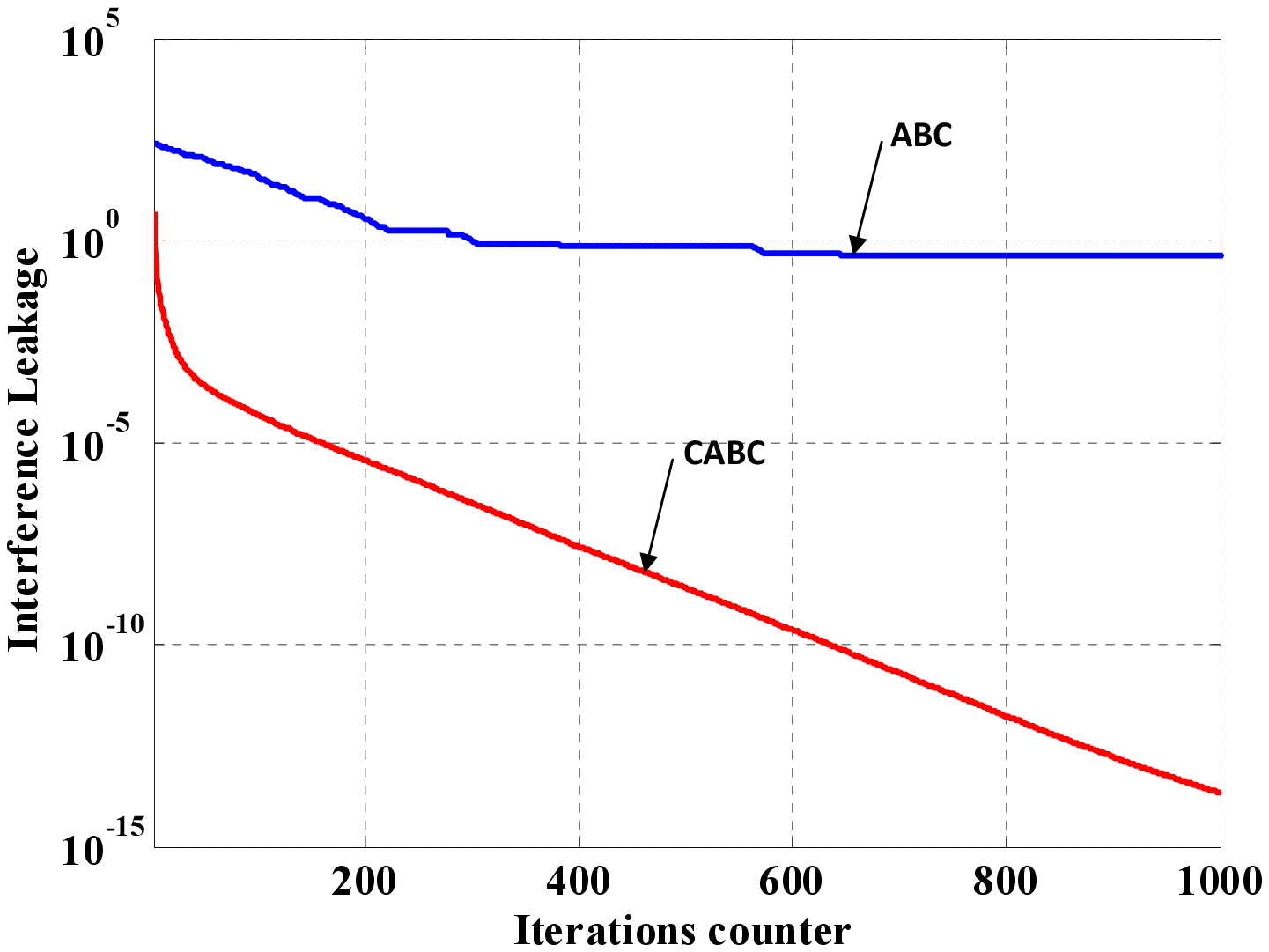}
  \caption{ABC and CABC simulation results for $\left(5 \times 5, 2\right)^{7}$ scenario.}
  \label{fig:fig7}
\end{flushleft}
\end{figure}

\begin{figure} 
\begin{center}
\includegraphics[width=0.4\textwidth]{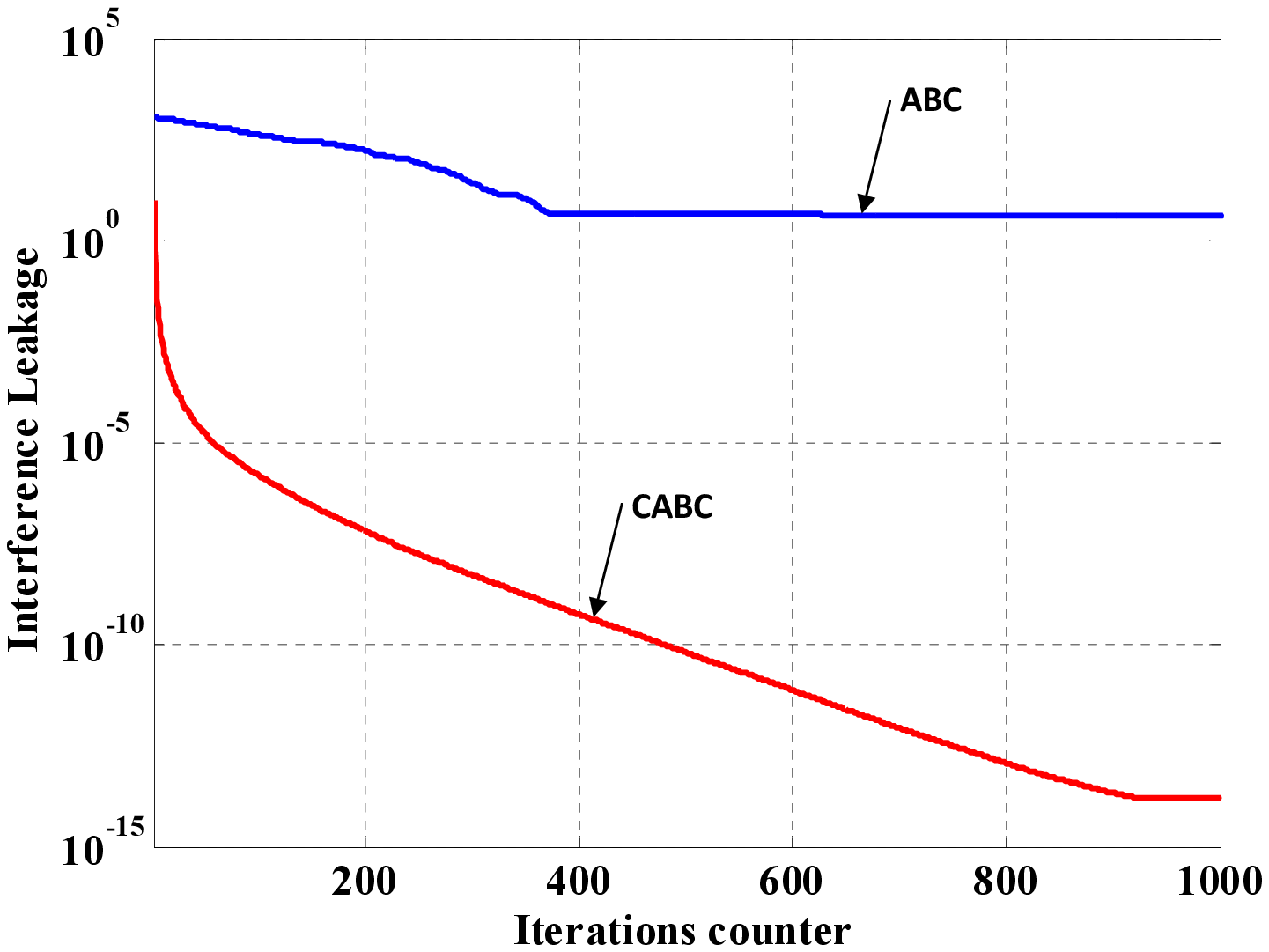}
  \caption{ABC and CABC simulation results for $\left(5 \times 5, 2\right)^{13}$ scenario.}
  \label{fig:fig13}
\end{center}
\end{figure}

The same scenarios were optimized using ABC and CABC. For ABC we used $SN=100$ and $Limit=5$, as for PSO the calculations are deliberately stopped after 1000 iterations after premature convergence. Concerning CABC, an equal number swarms that decision variables was optimized. Each swarm is 15 sized, $Limit$ is kept equal to 5. According to Table \ref{resultsABC}, the IL has reached very satisfactory values (the IL was divided by relatively $10^{14}$ given that the original purpose is to nullify it). The approach by CABC optimization roughly keeps the same performances for all the ascending values $K$. Obviously the CC framework has significantly improved the ABC algorithm performance when dealing with large scale optimizations procedures, also be noted the rapid convergence due to the small swarms size. Fig.\ref{fig:fig3} to Fig.\ref{fig:fig13} depict, for each scenario, the convergence plot, in CABC cases the IL level is fairly low for a signal-to-noise ratio (SNR) regime where IA is significant.

\begin{table}
\begin{flushleft}
\begin{tabular}{|l|l|l|}
\hline
Scenarios & Dimension & minimum IL (PSO algorithm)\\
\hline
$\left(5 \times 5, 2\right)^3$ & 120 &  0.0024\\
\hline
$\left(5 \times 5, 2\right)^7$ & 280 & 0.2479\\
\hline
$\left(5 \times 5, 2\right)^{13}$ & 520 & 2.4852\\
\hline \hline
Scenarios & \multicolumn{2}{|l|}{minimum IL (CPSO algorithm)}\\
\hline
$\left(5 \times 5, 2\right)^3$ &\multicolumn{2}{|l|}{$5.8298\times 10^{-5}$}\\
\hline
$\left(5 \times 5, 2\right)^7$ & \multicolumn{2}{|l|}{$6.6484\times 10^{-5}$}\\
\hline
$\left(5 \times 5, 2\right)^{13}$ & \multicolumn{2}{|l|}{$4.2649\times 10^{-6}$}\\
\hline
\end{tabular}
\end{flushleft}
\caption{PSO and CPSO simulation results with $\left(M \times N,d\right)^K$ representation}
\label{resultsPSO}
\end{table}

\begin{table}
\begin{flushleft}
\begin{tabular}{|l|l|l|}
\hline
Scenarios & Dimension & minimum IL (ABC algorithm)\\
\hline
$\left(5 \times 5, 2\right)^3$ & 120 &  0.0163\\
\hline
$\left(5 \times 5, 2\right)^7$ & 280 & 0.4079\\
\hline
$\left(5 \times 5, 2\right)^{13}$ & 520 & 4.2834\\
\hline \hline
Scenarios & \multicolumn{2}{|l|}{minimum IL (CABC algorithm)}\\
\hline
$\left(5 \times 5, 2\right)^3$ &\multicolumn{2}{|l|}{$1.9650\times 10^{-15}$}\\
\hline
$\left(5 \times 5, 2\right)^7$ & \multicolumn{2}{|l|}{$2.2132\times 10^{-14}$}\\
\hline
$\left(5 \times 5, 2\right)^{13}$ & \multicolumn{2}{|l|}{$1.5036\times 10^{-14}$}\\
\hline
\end{tabular}
\end{flushleft}
\caption{ABC and CABC simulation results with $\left(M \times N,d\right)^K$ representation}
\label{resultsABC}
\end{table}

\section{Conclusion}
This paper proposed a novel AI scheme for $K$-User MIMO IC. More precisely, the purpose is to find a more effective and straightforward solution to the problem of IL optimizing when achieving IA. For this, we applied the two most well-known metaheuristics, namely PSO and ABC. The CC approach was privileged since the optimization problem is of large scale. The comparative study that was conducted on growing size $K$-User MIMO IC showed the superiority of CC-based ABC algorithm. The CABC based IA solutions exhibit a very satisfactory convergence behavior. What encourages asserting that the metaheuristic based IA solutions can be a serious alternative to the algebraic ones, because convergence is almost verified and less constrained with hard algebraic assumptions. Especially since the current trend is in the direction of increasing the number of antennas in MIMO systems.
\bibliographystyle{plain}
\bibliography{Biblio_DAT}
\end{document}